\documentclass[10pt,twocolumn,twoside]{IEEEtran}
\usepackage{bm}
\usepackage{color}
\usepackage[linesnumbered,boxed,commentsnumbered,ruled,vlined,longend]{algorithm2e}
\usepackage{amsmath}
\usepackage{amssymb}
\usepackage{arydshln}
\usepackage{textcase}
\usepackage{soul}
\usepackage{indentfirst}
\usepackage{graphicx}
\usepackage{epsfig,epsf,color,amsmath,cite,epic,eepic}
\usepackage{cite}
\usepackage[colorlinks = true,
linkcolor = red,
urlcolor  = red,
citecolor = blue,
anchorcolor = blue]{hyperref}
\IEEEoverridecommandlockouts

\usepackage{enumerate}

\title{Tuning-Free, Low Memory Robust Estimator\\ to Mitigate GPS Spoofing Attacks}
\author{
	{Junhwan Lee, 
	Ahmad F. Taha,
	Nikolaos Gatsis, and David Akopian. } 
	\thanks{
		{The authors are all with the Department of Electrical and Computer Engineering at the University of Texas, San Antonio.  \{junhwan.lee,ahmad.taha,nikolaos.gatsis,david.akopian\}@utsa.edu. We acknowledge the financial support by the National Science Foundation under Grant ECCS-1719043, and the suggestions made by the editor and reviewers.  }}
	}
\begin{document}
		\maketitle	
	\begin{abstract}
The operation of critical infrastructures  such as the electrical power grid, cellphone towers, and financial institutions relies on precise timing provided by stationary GPS receivers. These GPS devices are vulnerable to a type of spoofing called Time Synchronization Attack (TSA), whose objective is to maliciously alter the timing provided by the GPS receiver.  The objective of this paper is to design a tuning-free, low memory robust estimator to mitigate such spoofing attacks. The contribution is that the proposed method dispenses with several limitations found in the existing state-of-the-art methods in the literature that require parameter tuning, availability of the statistical distributions of noise, real-time optimization, or heavy computations. Specifically, we \textit{(i)} utilize an observer design for linear systems under unknown inputs, \textit{(ii)}  adjust it to include a state-correction algorithm, \textit{(iii)} design a realistic experimental setup with real GPS data and sensible spoofing attacks, and \textit{(iv)} showcase how the proposed tuning-free, low memory robust estimator can combat TSAs. Numerical tests with real GPS data demonstrate that accurate time can be provided to the user under various attack conditions.  
		\end{abstract}
		

	\section{Introduction and Motivation}
	\IEEEPARstart{T}{he} Global Positioning System (GPS) is widely utilized in an abundance of applications. 
	The study~\cite{GAO} in particular emphasizes how critical infrastructures such as communications, the power grid, transportation, and even financial services can be disrupted if the integrity of the GPS is compromised. 
	
	 Since most systems rely on non-encrypted civilian GPS signals~\cite{Xu}, the GPS is vulnerable to intentional attacks. There are two types of deliberate attacks on GPS: jamming and spoofing~\cite{b2}. While jamming completely blocks signal reception by transmitting high power noise, spoofing changes the transmitted signal or data to deceive the GPS receiver. Various experiments have shown that different types of spoofing attacks such as data level spoofing, signal level spoofing, delaying, and record-and-reply attacks~\cite{b3,b5,b11} could affect off-the-shelf GPS receivers.

	  A Time Synchronization Attack (TSA) is a particular class of spoofing attacks on \emph{stationary} GPS receivers that provide precise timing in various applications, including phasor measurement units (PMUs), cellphone towers, and financial institutions~\cite{b1,b7}. The objective of the attack is to mislead the time estimated by the receiver which, in the case of the power grid for example, can disrupt the reliable monitoring of the grid's cyber-physical status. 
	 
Countermeasures against spoofing attacks have been proposed in~\cite{b10,6884182,8245836} and include techniques that rely on multiple GPS receivers~\cite{b8,b9} or check the magnitude of error in the GPS data~\cite{b10}; see~\cite{b5} for a review of anti-spoofing techniques. Another approach to mitigate and detect spoofing TSAs is through robust, dynamic state estimation routines that are designed to deal with outliers and malicious attacks.  
	 
	 The robust state estimation literature is indeed rich with two main classes of methods. The first class is based on robust observers and Lyapunov theory which often does not assume any statistical distribution for unknown inputs or noise \cite{bhattacharyya1978observer,kang2013survey}. The second class is based on Kalman filter and its derivatives, that often assume statistical distribution of noise~\cite{RobustSmoother,jazwinski2007stochastic,einicke1999robust}.  Relevant to the robust estimation problem under TSAs, a novel anti-spoofing particle filter is devised to find the receiver position even under spoofing interference~\cite{6884182}. In our recent work~\cite{8245836}, we develop a real-time optimization method to detect and mitigate TSAs using weighted $\ell_1$ minimization. The aforementioned methods all require either tuning of some parameters, real-time optimization, or availability of the statistical distribution of noise.

\textcolor{black}{The objective of this paper is to design a robust state estimator that combats TSAs while being endowed with the following properties: \textit{(i)} It is a tuning-free method that does not require any training; \textit{(ii)} it has low memory requirement in the sense no heavy computations are needed in real-time; and \textit{(iii)} the designed tuning-free, low-memory robust estimator can correctly reconstruct the actual physical state of the receiver using a  realistic testbed. We note here that our objective is not to develop a generalized theory for robust estimation, but rather to build on the recent theoretical advancement in this field and adapt it to the GPS spoofing problem through a realistic testbed. The robust state estimation method presented in this paper is an adaptation of the method in~\cite{charandabi2012observer}. The motivation for using this estimator is provided in great detail in the next sections; the paper's organization is given next.}

Section~\ref{sec:dynamics} presents the dynamic modeling of bias and drift in GPS receivers and showcases how TSA attacks can be modeled and designed to mislead standard state estimators. Section~\ref{sec:RSE} presents a robust state estimator in addition to a state correction and TSA reconstruction routine. Section~\ref{sec:numtests} concludes the paper with realistic numerical tests on real GPS data. For reproducibility, the data used in the numerical tests and the results are all provided through a Github link.

	\section{Dynamic Modeling Under GPS Spoofing}~\label{sec:dynamics}
 The primary goal of GPS localization is to accurately estimate the {p}osition, {v}elocity, clock bias, and clock drift of the receiver in every time step---conventionally referred to as the \textbf{p}osition, \textbf{v}elocity, and \textbf{t}ime (PVT) solution. 
	 
	 The position of the GPS receiver (user) in Earth Centered Earth Fixed coordinates is denoted by $\bm {p}_{\mathrm{u}}=[x_{\mathrm{u}},y_{\mathrm{u}},z_{\mathrm{u}}]^{\top}$. 
	 To estimate the receiver's location and velocity, the GPS exploits the known location of satellites and the distance from each satellite to the receiver. Let $N$ denote the number of fixed satellites visible by the receiver, then $\bm {p_{n}} = [x_{n}(t_{n}),y_{n}(t_{n}),z_{n}(t_{n})]^{\top}$ for $n = 1,2,\ldots,N$ is the satellite position at the time of transmission $t_{n}$. Also, we consider that $t_{R}$ models the arrival time of transmitted signal at the receiver. The approximate distance between each satellite and the receiver can be written as $\rho_{n} = c(t_{R}-t_{n})$, where $c$ is the speed of light and  $\rho_{n}$ is commonly known as the pseudorange~\cite{r27}. The pseudoranges differ from the true distances because $t_{R}$ and $t_{n}$ are offset by clock biases denoted by $b_u$ and $b_n$. This relationship is established as
	  \vspace{-5pt}
	  \begin{equation}
	  \label{clock}
	  \begin{split}
	  t_{R} = t^{\mathrm{GPS}}_{R} + b_\mathrm{u},\;\; t_{n} = t^{\mathrm{GPS}}_{n} + b_{n} 
	  \end{split}
	  \end{equation}
	  where $t^{\mathrm{GPS}}_R$ defines the reception time in the absolute GPS time, while $t^{\mathrm{GPS}}_n$ is the signal transmission time in GPS time. Then, if $d_{n}$ represents the true distance from the receiver to each satellite, it holds that $d_n = c(t^{\mathrm{GPS}}_{R}-t^{\mathrm{GPS}}_{n})$ using the unbiased transmission and reception times. Alternatively, $d_n$ can be expressed by taking the 2-norm of position difference between the satellite and the receiver, given by $d_{n} = ||\bm {p_{n}}-\bm {p_\mathrm{u}}||_{2}$. The following pseudorange equation is generated by combining the previous two equations for $d_n$~\cite{r27}
	\begin{equation}
	\label{Pseudorange}
	\begin{split}
	\rho_{n}=\Arrowvert \bm {p}_{n}- \bm {p}_{\mathrm{u}} \Arrowvert_2+c(b_{\mathrm{u}}-b_{n})+\epsilon_{\rho_{n}}
	\end{split}
	\end{equation}
	where $\epsilon_{\rho_{n}}$ captures atmospheric effects and receiver noise.  
	
	In addition to the pseudoranges, the GPS receiver can also measure the rate at which the pseudoranges vary over time, denoted by $\dot{\rho}_{n}$, and called pseudorange rate. 
	 The pseudorange rates are expressed in terms of the satellite velocities $\bm {v}_n$ and the user (GPS receiver) velocity $\bm {v}_\mathrm{u}$ as
	\vspace{-5pt}
	\begin{equation}
	\label{Pseudorange Rate}
	\begin{split}
	\dot{\rho}_{n}  = (\bm {v}_{n}-\bm {v}_{\mathrm{u}})^{\top}\frac{\bm {p}_{n}-\bm {p}_{\mathrm{u}}}{\Arrowvert \bm {p}_{n}-\bm {p}_{\mathrm{u}} \Arrowvert}+\dot{b}_{\mathrm{u}}
	+\epsilon_{\dot{\rho}_{n}}
	\end{split}
	\end{equation}  
	where $\dot{b}_{\mathrm{u}}$ represents the GPS receiver clock drift and  $\epsilon_{\dot{\rho}_{n}}$ is the noise. 
	 In~\eqref{Pseudorange} and \eqref{Pseudorange Rate}, the unknown PVT variables  ()user position ($\bm {p}_{\mathrm{u}}$), user velocity ($\bm {v}_{\mathrm{u}}$), clock bias ($b_{\mathrm{u}}$), and clock drift ($\dot{b}_{\mathrm{u}}$)] are usually computed using nonlinear weighted least squares.
	 
	\par The random walk model captures the dynamics relating variables in~\eqref{Pseudorange} and \eqref{Pseudorange Rate} for stationary applications~\cite{r27}. 
	The following is the stationary random walk model:
	\begin{equation}
	\label{randomwalkmodel}
	\begin{bmatrix}
	x_{\mathrm{u}} [k+1] \\ y_{\mathrm{u}}[k+1]  \\ z_{\mathrm{u}}[k+1]  \\ b_{\mathrm{u}}[k+1]  \\ \dot{b}_{\mathrm{u}}[k+1] 
	\end{bmatrix} = \begin{bmatrix}
	\begin{array}{c;{2pt/2pt}r}
	\bm {I}_{3 \times 3} & \begin{matrix} \bm {0}_{3 \times 2} \end{matrix}  \\ \hdashline[1.5pt/1.5pt]
	\bm {0}_{2 \times 3} & \begin{matrix} 1 & \Delta t \\ 0 & 1 \end{matrix}  
	\end{array}
	\end{bmatrix} \begin{bmatrix}
	x_{\mathrm{u}} [k] \\ y_{\mathrm{u}}[k]  \\ z_{\mathrm{u}}[k]  \\ b_{\mathrm{u}}[k]  \\ \dot{b}_{\mathrm{u}}[k] 
	\end{bmatrix}+ \bm {w}[k]
	\end{equation}
	where $k$ is the time index; $\Delta t$ is the time resolution; and $\bm {w}$ is noise in the system. Generally, stand-alone receivers like those present in PMUs, use the Extended Kalman Filter (EKF) to estimate the PVT solution~\cite{r27}. 
	
	\par Since the receiver is stationary, the position ($\bm {p}_\mathrm{u}$) can be treated as known constant while the receiver velocity ($\bm {v}_\mathrm{u}$) is known to be zero. Thus, the only variables to be estimated are in fact the clock bias and drift, ${b}_{\mathrm{u}}[k]$ and $\dot{b}_{\mathrm{u}}[k]$. 
	 Based on~\eqref{Pseudorange}, \eqref{Pseudorange Rate} and the dynamic model in \eqref{randomwalkmodel}, the fundamental plant model is constructed as follows using $\boldsymbol{\rho}[k] = [\rho_1[k], ... , \rho_N[k]]^{\top}$ and $\boldsymbol{\dot{\rho}}[k] = [\dot{\rho}_1[k], ... , \dot{\rho}_N[k]]^{\top}$:
	\begin{equation}
	\label{statedynamicalModel}
	\begin{split}
	{\begin{bmatrix}
		cb_{\mathrm{u}}[k+1]\\c\dot{b}_{\mathrm{u}}[k+1]
		\end{bmatrix}} = \bm {A}{\begin{bmatrix}
		cb_{\mathrm{u}}[k]\\c\dot{b}_{\mathrm{u}}[k] 
		\end{bmatrix}} + \bm {w}[k] \\
	\end{split}
	\end{equation}
	
	\begin{equation}
	\label{observationModel}
	\begin{split}
	\begin{bmatrix}
	\boldsymbol{\rho}[k]\\ \boldsymbol{\dot{\rho}}[k]
	\end{bmatrix} = \bm {C}{\begin{bmatrix}
		cb_{\mathrm{u}}[k]\\c\dot{b}_{\mathrm{u}}[k] 
		\end{bmatrix}} + \bm {c}_{l}[k] + \bm {\epsilon}[k]
	\end{split}
	\end{equation}
	where 
	\hspace{-5pt}
	\small
	\begin{description}
		\item{$\bm {A}$} = $\begin{bmatrix}
		1 & \Delta t\\ 0 & 1
		\end{bmatrix} $,\;	
		 {$\bm {C}$} = $\begin{bmatrix}
		\bm {1}_{N \times 1}&\bm {0}_{N \times 1}\\\bm {0}_{N \times 1} & \bm {1}_{N \times 1}
		\end{bmatrix}$ 
		\item {$\bm {c}_l[k]$} = $\begin{bmatrix} 
		\Arrowvert \bm {p}_{1}[k]-\bm {p}_{\mathrm{u}}[k]\Arrowvert -cb_{1}[k] \\\vdots\\\Arrowvert \bm {p}_{N}[k]-\bm {p}_{\mathrm{u}}[k]\Arrowvert - cb_{N}[k]\\
		(\bm {v}_{1}[k]-\bm {v}_{\mathrm{u}}[k])^{\top}.\dfrac{\bm {p}_{1}[k]-\bm {p}_{\mathrm{u}}[k]}{\Arrowvert \bm {p}_{1}[k]-\bm {p}_{\mathrm{u}}[k] \Arrowvert} - c\dot{b}_{1}[k]\\ \vdots \\ (\bm {v}_{N}[k]-\bm {v}_{\mathrm{u}}[k])^{\top}.\dfrac{\bm {p}_{N}[k]-\bm {p}_{\mathrm{u}}[k]}{\Arrowvert \bm {p}_{N}[k]-\bm {p}_{\mathrm{u}}[k] \Arrowvert} -c\dot{b}_{N}[k]
		\end{bmatrix}$
	\end{description}
	\normalsize
	and $\bm {w}[k]$ and $\bm {\epsilon}[k]$ represent process/measurement noise; vector $\bm {c}_l[k]$ is based on the known satellite position, velocity and clock characteristics---a time-varying, known quantity. Equations~\eqref{statedynamicalModel} and \eqref{observationModel} can be written as
	\begin{equation}
	\label{StateSpaceModel}
	\begin{split}
	\bm {x}[k+1]& = \bm {A}\bm {x}[k] + \bm {w}[k] \\
	\bm {y}[k] &= \bm {C}\bm {x}[k] + \bm {c}_l[k] + \bm {\epsilon}[k].
	\end{split}
	\end{equation}
	
	 The state space model \eqref{StateSpaceModel}, however, does not model potential spoofing attacks. \textcolor{black}{While many different physical spoofing mechanisms are devised to deceive the victim receiver~\cite{b11}, time synchronization attack (TSA) is applied on the stationary GPS receiver. In practical sense, TSA alters the timestamp estimate by inserting the spoofing signal into the authentic pseudorange signals:} 
	$$ \textcolor{black}{\boldsymbol{\rho_{s}[k] = \boldsymbol{\rho}[k] + s_{\rho}[k] } ,\; \boldsymbol{\dot{\rho}_{s}[k] = \boldsymbol{\dot{\rho}}[k] + s_{\dot{\rho}}[k] }. }$$
	 where $\boldsymbol{s_{\rho}}[k]$ and $\boldsymbol{s_{\dot{\rho}}}[k]$ denote the spoofing attacks, and $\boldsymbol{\rho}_{s}[k]$ and $\boldsymbol{\dot{\rho}_{s}}[k]$ are the spoofed measurements. 
	 
	 \textcolor{black}{Specifically, there are two different types of TSAs according to the shape of $\boldsymbol{s_{\rho}}$. While Type I attack injects an abrupt signal, e.g., $\boldsymbol{s_{\rho}}[k>\alpha] = 8000\ \mathrm{m}$ where $\alpha$ indicates the initial time of attack, Type II attack modifies the clock bias in gradual manner manipulated by $\boldsymbol{s_{\rho}}[k] = \boldsymbol{s_{\rho}}[k-1]+ \boldsymbol{s_{\dot{\rho}}}[k]\Delta t$; see~\cite{Shepard2012}}. The actual effect of each type of attack on the clock state is thoroughly reviewed in~\cite{b12}. 
	 
	 As an example, in order for the spoofing signal to be considered as intentional attack on a PMU, it has to satisfy certain conditions. According to the IEEE C37.118 Standard, the attack has to result in 1\% total variation error, which is equivalent to 26.65 $\mu$s clock bias error, or 7989 $\mathrm{m}$ of distance equivalent bias error in order for the attack to be considered infringing~\cite{IEEEStandard}.
	 These types of spoofing attacks---regardless of their physical mechanism---impact the state dynamics as $$ \bm x[k+1] = \bm A \bm x [k] +\bm d[k]$$
	where $\bm d[k] = [d_1[k] \ d_2[k]]^{\top}$ models and lumps TSAs and any process noise. Concrete examples of TSAs are given in Section \ref{sec:RSE}. In our previous work~\cite[Section III]{8245836}, we show how specific forms of $\bm d[k]$ can mislead the receiver.
	\section{Robust State Estimator}~\label{sec:RSE}
	In this section, we present a state estimation algorithm that is endowed with the following properties: \textit{(i)} It is a tuning-free method that does not require any knowledge of noise distribution, initial parameters or states, or other coefficients; \textit{(ii)} it has low memory requirements in the sense no heavy computations are needed which is befitting to devices with limited computational power and limited internet connectivity; \textit{(iii)} it is robust to GPS spoofing, time-synchronization attacks.  
	\subsection{GPS Clock Model and  Estimator Dynamics} The plant model under the spoofing attack based on the previous can be written as
	\vspace{-5pt}
	\begin{equation}
	\label{Plant Model}
	\begin{split}
	\bm x[k+1] = &\bm A\bm x[k]+\bm d[k]\\
	\bm 	y[k] = &\bm C\bm x[k]+\bm c_{l}[k] + \bm \epsilon[k]\\
	\end{split}
	\end{equation}
	where $\bm x[k] \in \mathbb{R}^{2}$ represents the state vector of clock bias and drift at time $k$; $\bm y[k] \in\mathbb{R}^{2N}$ represents a single column vector of pseudoranges and pseudorange rates where $N$ indicates the fixed number of visible satellites at every time index; 
$\bm d[k]\in \mathbb{R}^{2}$ is the unknown spoofing attack applied to the bias and drift which also includes process noise; state-space matrices $\bm A$, $\bm C$, $\bm c_{l}$ are discussed in the previous section. 
	
	Consider now a new \textit{{m}odified} state vector $\bm x_m[k] \in \mathbb{R}^2$ which represents the state vector without spoofing attack and follows the following dynamics:
	\begin{equation}
	\label{New Variable}
	\begin{split}
	\bm x_m[k+1] = \bm x[k+1]-\bm d[k] = \bm A\bm x[k].
	\end{split}
	\end{equation}
	The left-hand side of~(\ref{New Variable}) essentially represents the original state vector considering that the spoofing attack $\bm d[k]$ is removed. The modified state vector $\bm x_m[k]$ propagates through to $\bm y[k]$. This yields:
	\begin{equation}
	\label{New Modified Plant}
	\begin{split}
	\bm x_m[k+1] = &\bm A\bm x_m[k]+\bm A\bm d[k-1] \\
	\bm y[k] = &\bm C\bm x_m[k]+\bm c_{l}[k]+\bm C \bm d[k-1]. \\
	\end{split}
	\end{equation}
	The presented state estimator in this paper is an adaptation of the observer from~\cite{charandabi2012observer} and follows the difference equation:
	\begin{align}
	& \textsc{\textbf{RobustEstimator}} \notag \\
\hspace{-0.6cm}	\hat{\bm	x}_{m}[k+1] = & \bm	A\hat{\bm x}_m[k]+\bm	A\hat{\bm	d}[k-1] + \bm	L_1(\bm	y[k]-\hat{\bm	y}[k])  \notag \\
	\bm	\hat{\bm	y}[k] = &\bm	C\hat{\bm	x}_{m}[k]+\bm	c_{l}[k]+\bm	C\hat{\bm	d}[k-1] \notag  \\
\bm	e[k]=&\bm y[k] - \hat{\bm y}[k] \notag \\
	\hat{\bm {d}}[k] =& \hat{\bm {d}}[k-1] + \bm {L_2} \bm {C}^{\top}\bm {e}[k]\label{Modified Observer Design}
	\end{align}
	where $\hat{\bm x}_{m}[k] \in \mathbb{R}^2$ is a state estimate of corrected state vector ${\bm x}_{m}[k]$ at time $k$; $\hat{\bm y}[k]\in \mathbb{R}^{2N}$ is the estimate vector of observation $\bm y[k]$; $\hat{\bm d}[k]$ is an estimate of the spoofing attack $\bm d[k]$. {We note here that $\hat{\bm d}[-1], \hat{\bm x}_m[0]$, and $\hat{\bm y}[0] $ should be initialized before iteration starts at $k=0$ with arbitrary initial conditions.} Matrices $\bm L_1\in \mathbb{R}^{2\times 2N}$ and $\bm L_2 \in \mathbb{R}^{2\times 2}$ are optimization variables where $\bm L_1$ is akin, in principle, to a Luenberger gain that is designed here to ensure robustness of the state estimation to spoofing attacks. 
	\subsection{Design of Robust Gains $\textbf{\textit{L}}_{1,2}$}
	The design of the robust estimator gains $\bm L_{1,2}$ is based on linear matrix inequalities (LMIs). Simply put, the objective of the designed observer is to guarantee asymptotically stable estimation error dynamics. That is, matrices $\bm L_1$ and $\bm L_2$ are designed to guarantee that $\lim_{k\rightarrow \infty} \bm e[k] = 0$ under non-zero spoofing attack $\bm d[k]$ and bounded estimation error under spoofing attacks.  The \textsc{\textbf{RobustEstimator}} variables are designed via solving this low-dimensional feasibility problem with one linear matrix inequality (LMI), given as follows:
	\begin{subequations}	\label{gainLMI}
		\textsc{\textbf{EstimatorDesign}} \begin{align}
		%
\hspace{-1.5cm}		\mathrm{find} \;&\bm G \in \mathbb{R}^{2\times 2N},  \bm P\in \mathbb{R}^{2\times 2}, \notag  \bm Q\in \mathbb{R}^{2N\times 2N}, \bm M \in \mathbb{R}^{2\times 2}\\
		\mathrm{s.t.}
		&	{ \begin{bmatrix}
			\bm P & \star & \star & \star \\
			\bm 0 & \bm Q & \star &\star \\
			\bm G\bm C-	\bm P \bm A	& \bm G\bm C -\bm P \bm A& \bm P & \star \\
			\bm M \bm C^{\top} \bm C& \bm M \bm C^{\top} \bm C -\bm Q & \bm 0 & \bm Q
			\end{bmatrix} \succ \bm 0 	}\\
		&  \{\bm P, \bm Q, \bm M\}= \{\bm P^{\top}, \bm Q^{\top}, \bm M^{\top}\} \succ \bm 0,
		\end{align} 
	\end{subequations}
	where the symbol $\star$ is used to represent symmetric components in symmetric block matrices.
	After solving \eqref{gainLMI} for positive definite matrix variables $\bm P, \bm Q,$ and $\bm M$, and real matrix variable $\bm G$, the observer gains are computed as follows
	\begin{equation}
	\label{calculateGains}
	\bm L_1 = \bm P^{-1}\bm G ,\;\; \bm L_2 = \bm M\bm Q^{-1}.
	\end{equation} 
As mentioned earlier, the state estimator design is derived from~\cite{charandabi2012observer}; the reader is referred to that paper for the derivation of the above LMIs. \textcolor{black}{Note that no tuning is required to solve \textsc{\textbf{EstimatorDesign}}, and this LMI can be solved analytically via evaluating the Karush-Kuhn-Tucker conditions for feasibility~\cite{boyd2004convex}. Furthermore, any convex optimization toolbox or LMI solver can be used to solve~\eqref{gainLMI}. These include Matlab's LMI solver,  CVX~\cite{grant2014cvx}, and Yalmip~\cite{lofberg2004yalmip}. } 

 We note the following. First, the necessary conditions for existence of $\bm L_{1,2}$ are standard. These conditions are \textit{(a)} the detectability of $(\bm A, \bm C)$; \textit{(b)} the classical rank matching condition stating that $\mathrm{rank}(\bm C \bm B_d) = \mathrm{rank}(\bm B_d)$ where $\bm B_d=\bm I_2$ is the matrix coefficient of $\bm d[k]$ in~\eqref{Plant Model}; and  \textit{(c)} bounded variations of the unknown spoofing signal $\bm d[k]$.  Second, this robust estimator not only estimates $\bm x[k]$, but also the spoofing attack $\bm d[k]$. This is instrumental in mitigating and correcting the attack.  The next section showcases the design of the gain matrices. 
\vspace{-0.43cm}

	\subsection{State Correction Algorithm under Spoofing}
	Upon solving the LMIs and running the \textsc{\textbf{RobustEstimator}} from any arbitrary initial conditions, the estimator is guaranteed theoretically to produce bounded estimation error $\bm e[k]$, thereby estimating the then-spoofed bias and drift, and reconstructing the spoofing attack $\hat{\bm d}[k]$. With that in mind, this does not indicate that the bias and drift are correctly estimated seeing that spoofing attack had already changed the state through the state propagation and difference equation. To that end, this section develops a state correction algorithm to recover the authentic bias and drift values in real-time. 

To that end,  the dynamics~\eqref{New Variable}  can be written as
	\begin{equation}
	\small
	\label{disturbanceError}
	\begin{split}
	\bm {x}_m[k+1] &= \bm {A}\bm {x}_m[k]+\bm {A}\bm {d}[k-1] \\
	&= \bm {A}(\bm {A}\bm {x}_m[k-1]+\bm {A}\bm {d}[k-2])+\bm {A}\bm {d}[k-1].
	\end{split}
	\end{equation}
	
	\begin{algorithm}[t]
	\caption{\text{Robust Bias and Drift Estimation}}\label{alg:RSE}
	\DontPrintSemicolon
	\textbf{input:} Number of satellites $N$,  matrices $\bm A$ and $\bm C$\;
	\textbf{initialize:} $\hat{\bm d}[-1], \hat{\bm x}_m[0], \hat{\bm y}[0] $ \;
	\textcolor{black}{\textbf{\textit{Offline Computations}}}\;
	Compute $\bm {G,P,Q,M}$ given $\bm A$ and $\bm C$ by solving \eqref{gainLMI}\;
	Obtain $\bm L_1$ and $\bm L_2$ from~\eqref{calculateGains}\;
	\textcolor{black}{\textbf{\textit{Online Computations}}}\;
	\textcolor{black}{\While{$k\geq 0$}{
			Obtain $\bm c_l[k]$ from satellite measurements\;
			Run~\eqref{Modified Observer Design} and obtain $\hat{\bm x}_m[k+1]$ and $\hat{\bm d}[k]$\;
			Perform spoofing attack correction \eqref{dhatcorrection}\;
			Perform state and output correction~\eqref{correction}\;
			$k\leftarrow k+1$ \;
		}
	}
	\textbf{output:} Attack-free $\hat{\bm x}_c[k]$, TSA estimates $\hat{\bm d}_c[k]$
\end{algorithm}

	This relationship reveals that current spoofing during one time instant comprises the cumulative attacks from the previous time steps. Consequently, the disturbance estimate $\hat{\bm {d}}$ is not sufficient enough to correct the attacked states. Rather, a new disturbance estimate vector $\bm {d}_{c}$ is formulated to account for the accumulated disturbances. Considering the estimate of spoofing attacks $\hat{\bm {d}}[k] = \begin{bmatrix}
	\hat{d}_1[k] & \hat{d}_2[k] \end{bmatrix}^{\top}$  computed by \eqref{Modified Observer Design}, we propose estimating the new disturbance estimate vector $\bm {d}_{c}$ via 
	\begin{equation}
	\label{dhatcorrection}
\hspace{-0.3cm}	
	\hat{\bm {d}}_c[k] =\begin{bmatrix}
	\displaystyle \sum_{l=1}^{k} \hat{d}_1[l] + 
	\displaystyle \sum_{l=1}^{k-1}(k-l)\hat{d}_2[l]\\ \displaystyle \sum_{l=1}^{k-1} \hat{d}_2[l] 
	\end{bmatrix}
	\end{equation}
	\normalsize
	 This equation acknowledges the fact that estimated state $\bm {x}_m$ is still contaminated by the attack from the past time step. Therefore, the corrected state $\bm {x}_c$ and authentic observation state $\bm {y}_c$ could be retrieved by subtracting $\hat{\bm {d}}_c$ from $\bm {x}_m$ and $\hat{\bm {y}}$ as follows:
	\begin{eqnarray}
	\label{correction}
	\begin{split}
	\hat{\bm {x}}_c[k] &= \hat{\bm {x}}_m[k] - \hat{\bm {d}}_c[k],\;\;
	\hat{\bm {y}}_c[k] = \hat{\bm {y}}[k] - \bm {C}\hat{\bm {d}}_c[k]
	\end{split}
	\end{eqnarray}
	
	Algorithm \ref{alg:RSE} showcases the overall problem design, robust state estimation, and the reconstruction of the corrected bias and drift of the GPS receiver. The algorithm takes as inputs: the fixed number of satellites $N$, $\bm A$ and $\bm C$,  and satellites data which is encoded through $\bm c_l[k]$. The algorithm is divided into two stages---an offline stage and an online one. In the offline stage, the  \textsc{\textbf{RobustEstimator}} gains $\bm L_1$ and $\bm L_2$ are computed via solving~\eqref{Modified Observer Design} and evaluating~\eqref{calculateGains}. The online stage includes running the  \textsc{\textbf{RobustEstimator}}~\eqref{Modified Observer Design} and the cumulative corrections for the states and spoofing attacks. The algorithm returns the attack-free, yet still slightly noisy estimates of the bias and drift $\hat{\bm x}_c[k]$ and an estimate of the actual spoofing attack $\hat{\bm d}_c[k]$. 
	
\textcolor{black}{It is noteworthy to mention the following. First, the offline component of the algorithm---albeit offline---can be solved analytically seeing that the problem dimension is very small, when considering that only few satellite measurements are needed. Second, the algorithm and the LMI feasibility problem both only require a fixed \textit{number} of satellites, rather than a fixed satellite combination. This is important considering that different satellites are visible each time. In short, the proposed algorithm in this paper only assumes a minimum fixed number of satellites $N$, where these $N$ satellites can be changing in real time without impacting the algorithm or the design of the robust estimator. }

\textcolor{black}{	Third, Algorithm~\ref{alg:RSE} works for any reasonable initial conditions, that is, the estimation should converge regardless of the initial conditions choice. Fourth, this method is truly tuning-free: no prior knowledge of the statistical distribution of noise, or prior knowledge or tuning of any parameters is needed. The algorithm is also low-memory, as the only computation needed to be performed online is running the \textsc{\textbf{RobustEstimator}} and the correction models---both require a small number of matrix-vector multiplications. This implies that the proposed algorithm can be implemented in low-memory devices without the need for any intensive computational effort or internet connectivity. Finally, we note that the proposed algorithm has no stopping criterion seeing that it runs in real time.}
	\section{Case Studies: A Realistic Testbed}~\label{sec:numtests}		
	This section discusses the detection and mitigation of TSAs via various approaches.  {First, the experimental procedure is discussed.} Then, we compare the performance of the extended Kalman filter (EKF)---which has long been used in the literature~\cite{r27} as a \textit{ground truth} for estimating the bias and drift---and the classical Luenberger observer under spoofing attacks. Then, the performance of the proposed robust estimator under TSAs is showcased, followed by thorough comparison of the performance of the approaches. \textcolor{black}{The following link includes all codes and data used to generate the results, including the acquired GPS data:} \textcolor{black}{\texttt{github.com/junhwanlee95/Robust-Estimator}.
	Table~\ref{table2} summarizes the important vector nomenclature. }

	{ 	\begin{table}[t]
		\caption{Vector nomenclature.}
		\label{table2}
		\vspace{-0.5cm}
	\begin{center}
			\begin{tabular}{|c|c|}
				\hline
				Notation & Description \\
				\hline
				\hline
 				$\bm {x}_{\mathrm{GT}}$ & ground truth state vector  \\
				$\bm {x}_{\mathrm{EKF}}$ & estimated state vector from EKF \\
				$\bm {x}_{\mathrm{Luen}}$ & estimated state vector from Luenberger Observer \\ 
				$\hat{\bm {x}}_m$ & modified state vector estimates \\
				$\hat{\bm {x}}_c$ & corrected state vector via \eqref{correction} \\ 
				$\bm {d}$ & attack vector vector applied to $\bm {x}$\\			
				$\hat{\bm {d}}_c$ & corrected spoofing attack vector via \eqref{dhatcorrection} \\
				\hline  
			\end{tabular}
		\end{center}
		\vspace{-0.65cm}	
	\end{table}}
		\vspace{-0.274cm}
	\subsection{Setup: Model Simulation \& Obtaining Raw GPS Data}
		A Google Nexus 9 tablet, which has an embedded GPS chipset, is used to collect real GPS signals, which are recorded on November 4, 2018 at the University of Texas at San Antonio main campus. The data are available for the reader through the aforementioned Github link. While the receiver acquires the signal, the device remained still to simulate the stationary scenario. Raw GPS data is post-processed to obtain pseudorange and pseudorange rate data by \textit{GNSS Logger}, the Android application released by the Google Android location team~\cite{GoogleSource}. Then, Type I and II attacks are injected into the pseudorange and pseudorange rate data to simulate spoofing as discussed in Section~\ref{sec:dynamics} and shown in~\cite[Section III]{8245836}. The initial conditions for the robust estimator are chosen to be different than the actual, ground truth conditions. 
	\subsection{EKF and Luenberger Performances Under Attacks}
\begin{figure}[t]
		\centering \includegraphics[scale=0.435]{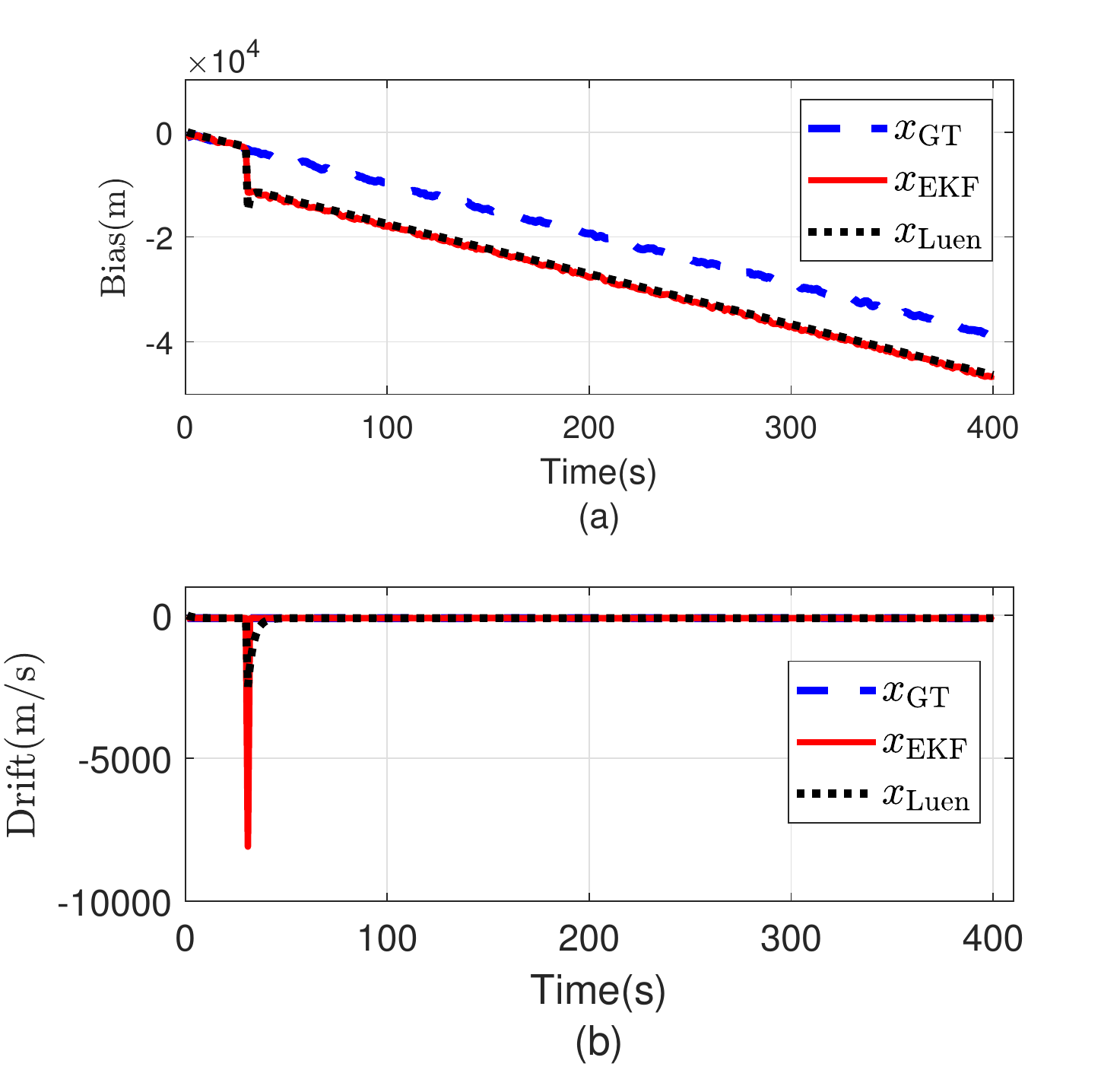}
		\vspace{-0.3cm}
		\caption{Comparison of ground truth, EKF and Luenberger observer estimates under Type I attack: (a) clock bias; (b) clock drift.}
		\label{ComparisonI}
		\vspace{-0.35cm} \end{figure}

\begin{figure}[t]
		\centering \includegraphics[scale=0.435]{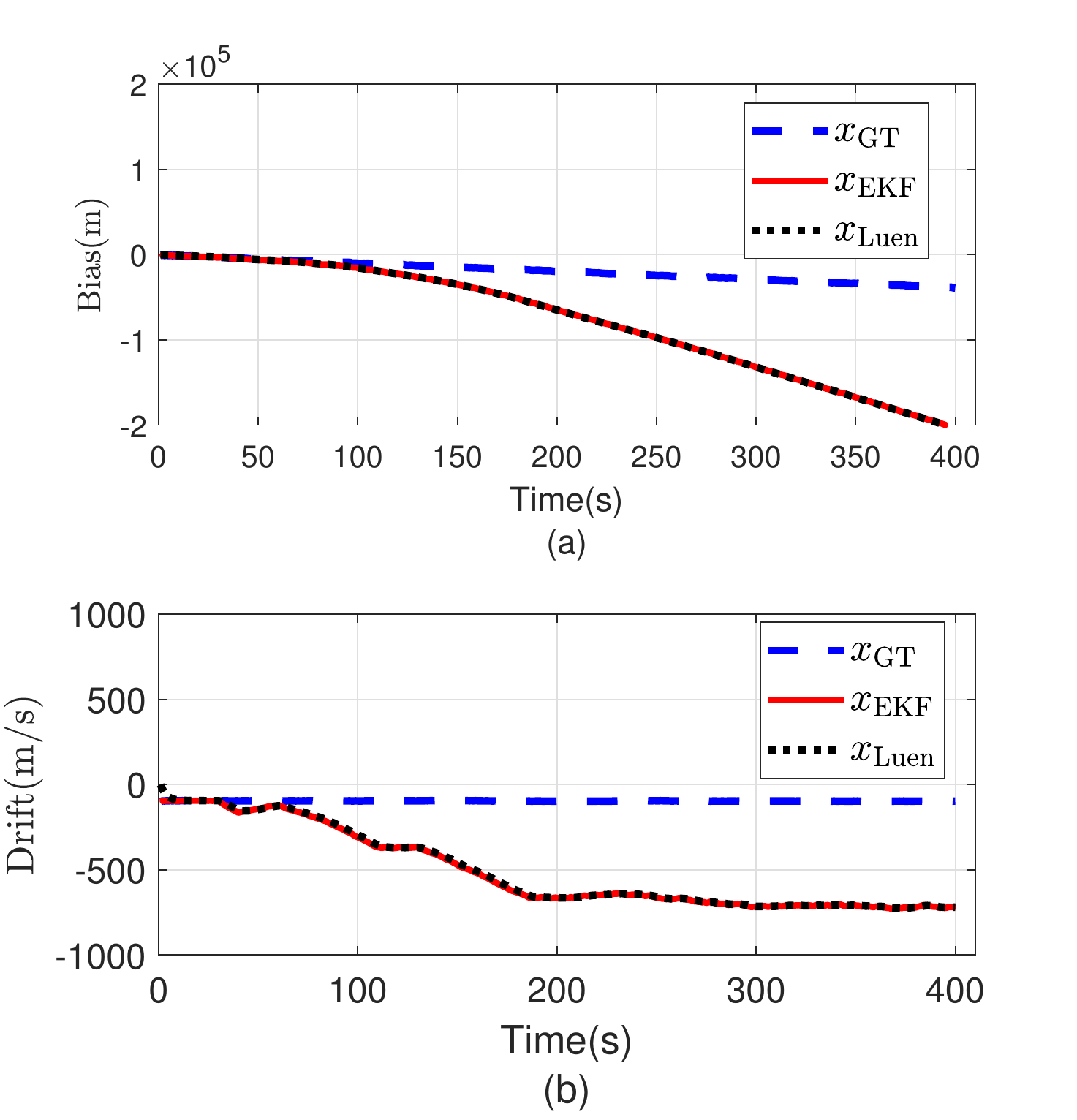}
		\vspace{-0.5cm}	\caption{Comparison of ground truth, EKF and Luenberger observer estimates under Type II attack: (a) clock bias; (b) clock drift.}
		\label{ComparisonII}
		\vspace{-0.55cm}
	\end{figure}

	\textcolor{black}{Here, we are interested in testing whether the EKF and the classical Luenberger observer~\cite{luenberger1966observers} can withstand TSAs of Type I and II. After Type I and II attacks are applied from $t=30 \ \mathrm{s}$ to $t=400 \ \mathrm{s}$, the performance of the EKF and Luenberger observer---which does not assume any statistical distribution about the noise---are shown in Fig.~\ref{ComparisonI} and~\ref{ComparisonII}.}

	\textcolor{black}{While the ground truth clock bias and drift, $\bm x_\mathrm{GT}$, are acquired through EKF by processing the authentic pseudorange data, the $\bm x_\mathrm{EKF}$ bias and drift are generated by applying the EKF to the spoofed pseudoranges. Another comparison to $\bm x_\mathrm{GT}$ is offered by the Luenberger observer estimate $\bm x_\mathrm{Luen}$ produced after designing the observer gain $\bm L_1$ such that the closed loop system eigenvalues are at $0.5$ and $0.7$.}
	\textcolor{black}{The performances of EKF and Luenberger observer are shown respectively  in Fig.~\ref{ComparisonI} and Fig.~\ref{ComparisonII}. It is evident that both approaches fail to estimate the correct states in the presence of the attack.  }
	
	\subsection{Robust Estimator Performance under Type I/II TSAs }
	
\begin{figure}[t]
		\centering\includegraphics[scale=0.28]{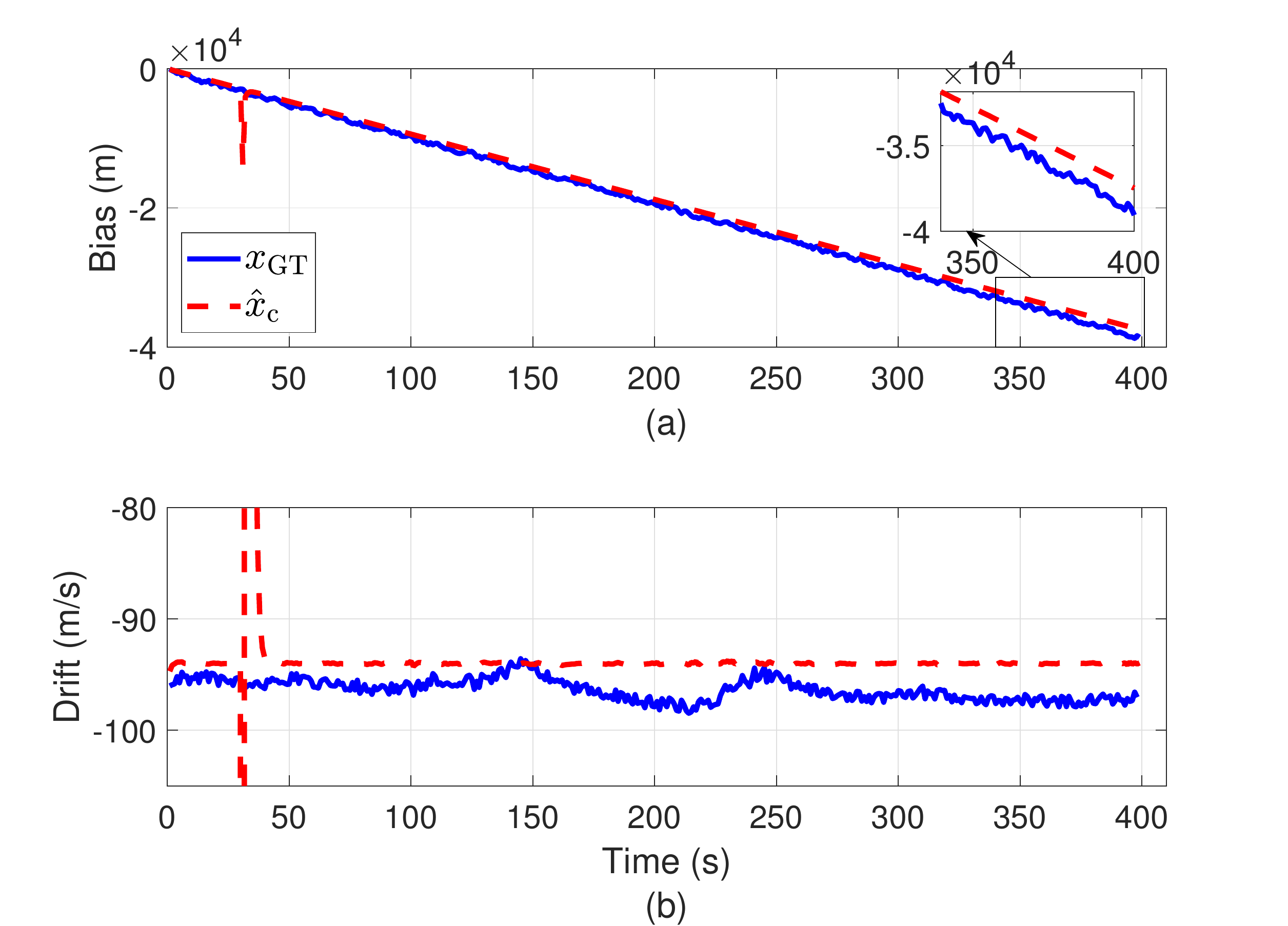}
				\vspace{-0.35cm}	
		\caption{Comparison of ground truth and corrected state through the  correction~\eqref{correction} under Type I attack: (a) clock bias; (b) clock drift.}
				\vspace{-0.35cm}	
		\label{CleanI}
	\end{figure}

	In this section, we test the proposed robust estimator and run Algorithm~\ref{alg:RSE} which contains an offline stage and an online one. First, we set $N=4$, i.e., we choose to sample data from only four satellites. \textcolor{black}{We solve the LMIs~\eqref{gainLMI} for the estimator gains.} 
	Using these estimator gains, the online portion of Algorithm~\ref{alg:RSE} is run. 
	\textcolor{black}{Pertaining to Type I attack, Fig.~\ref{CleanI} showcases the performance of \textsc{\textbf{RobustEstimator}}. Due to the attack at $t=30$ s, the $\hat{\bm x}_\mathrm{c}$ clock estimates are initially not correct, 
	but the clock bias and drift approach the respective ground truth values within approximately 3 and 11 seconds (cf. Fig.~\ref{CleanI}(a) and~\ref{CleanI}(b) respectively).}

\begin{figure}[t]
		\centering\includegraphics[scale=0.28]{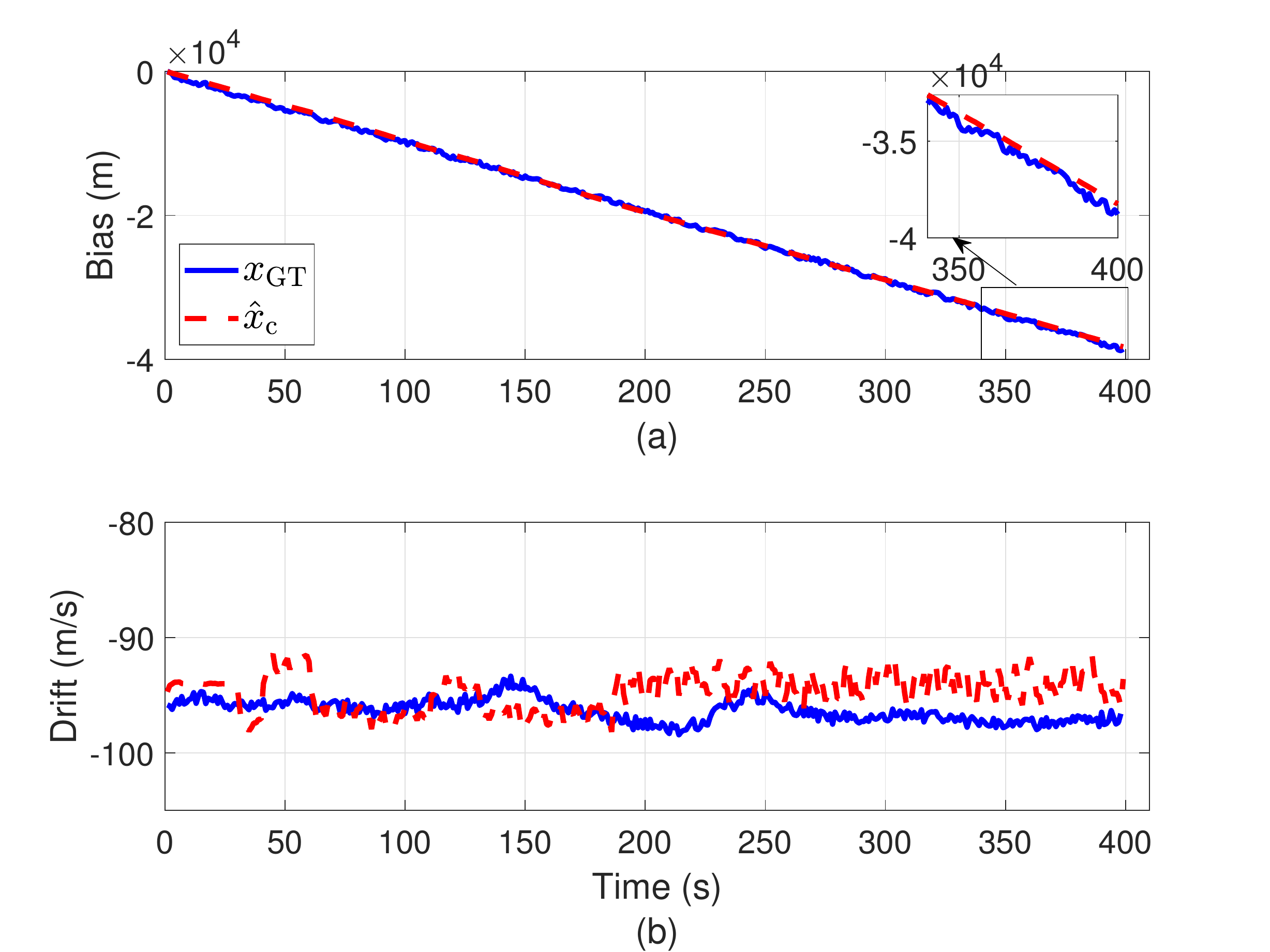}
		\vspace{-0.35cm}	\caption{Comparison of ground truth and corrected state through the correction~\eqref{correction} under Type II attack: (a) clock bias; (b) clock drift.}
				\vspace{-0.235cm}	
		\label{CleanII}
	\end{figure}

\begin{figure}[t]
	\includegraphics[scale=0.3]{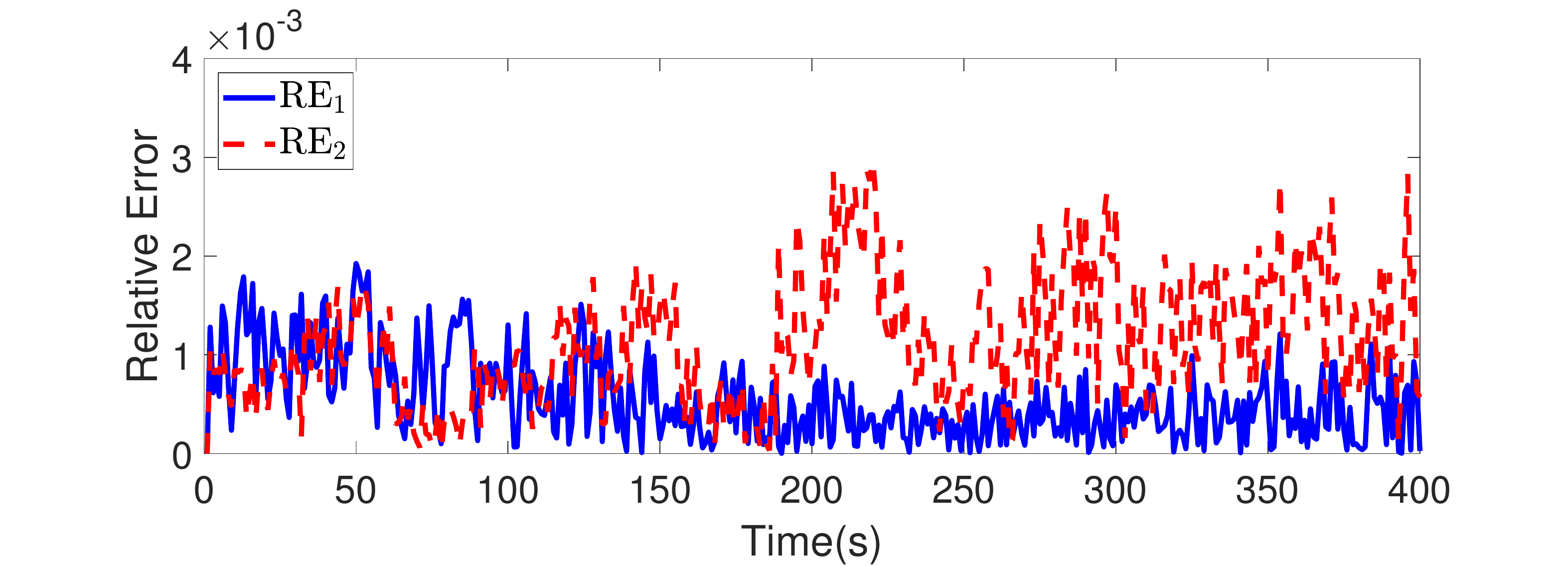}	
		\caption{Relative estimation error for clock bias and drift defined as: $\mathrm{RE}_1 = \frac{\mid\hat{\bm {x}}_{\mathrm{c}}(1)-\bm{x}_{\mathrm{GT}}(1)\mid}{\bm {x}_{\mathrm{GT}}(1)},$ $\mathrm{RE}_2 = \frac{\mid\hat{\bm {x}}_{\mathrm{c}}(2)-\bm{x}_{\mathrm{GT}}(2)\mid}{\bm {x}_{\mathrm{GT}}(2)}$.}
		\label{EstimationError}
				\vspace{-0.5cm}	
	\end{figure}

 	\textcolor{black}{Under the same condition and procedure, Algorithm \ref{alg:RSE} is applied on Type II attack to detect and correct the spoofed states. The results of obtaining the corrected state $\hat{\bm {x}}_c$ are shown in Fig.~\ref{CleanII}. In order to accurately depict the performance of \textsc{\textbf{RobustEstimator}}, the relative estimation error is calculated as $\frac{\mid\hat{\bm {x}}_{\mathrm{c}}-\bm {x}_{\mathrm{GT}}\mid}{{\bm {x}}_{\mathrm{GT}}}$ for each of the two states in $\bm x$ over time. The resulting graphs are shown in Fig.~\ref{EstimationError}. Comparison between $\hat{\bm {x}}_c$ and the $\bm x_\mathrm{GT}$ reveals that the maximum error between the two biases is $952.09\; \mathrm{m}$ or $3.17 \ \mu \mathrm{s}$. It is thus demonstrated that the Robust Estimator of Algorithm~\ref{alg:RSE} successfully detects and corrects the Type II attack.}
 
 
	In the interest of gauging the performance of  each approach, the root mean square error ($\mathsf{RMSE}$) of the estimated clock bias is calculated under both attack types. Let $K$ denote the total length of observation time ($K = 400$ in this experiment). The RMSE is defined as
	$
	\small
	\mathsf{RMSE}= \sqrt{\frac{1}{K}\sum_{k=0}^{K-1}  (c\hat{b}_{u}[k]-c\check{b}_{u}[k])^{2}}
	$
	where $c\check{b}_{u}[k]$ is the ground truth clock bias under normal conditions, and $c\hat{b}_{u}[k]$ equals to the estimated clock bias value from each approach. \textcolor{black}{Under Type II attack, the RMSE for EKF and the Luenberger observer are $\mathsf{RMSE}_{\mathrm{EKF}} = \mathrm{74344 \ m} $ and $\mathsf{RMSE}_{\mathrm{Luenberger}} = \mathrm{74433 \ m} $ respectively, while that of the robust estimator is $\mathsf{RMSE}_{\mathrm{RobustEstimator}} = \mathrm{354.9 \ m}$. As for Type I attack, the RMSEs are as follows: $\mathsf{RMSE}_{\mathrm{EKF}} = \mathrm{8477.3 \ m} $, $\mathsf{RMSE}_{\mathrm{Luenberger}} = \mathrm{8104.9 \ m} $, and $\mathsf{RMSE}_{\mathrm{RobustEstimator}} = \mathrm{1029 \ m}$.		} This illustrates the performance of this tuning-free, low-memory robust estimator in detecting spoofing attacks, while correctly reconstructing the bias and drift states of the GPS receiver. 
	

	\section{\textcolor{black}{Paper Summary and Future Work}}
	
	\textcolor{black}{In this paper, the design and realistic application of a low-memory, real-time \textsc{\textbf{RobustEstimator}} is studied. Utilizing the GPS receiver on a Google Nexus 9, real GPS data are collected and post-processed by injecting time-synchronization attacks to spoof the clock bias and drift of the device. Two types of attacks are introduced, and tested by the designed estimator. The  estimator successfully detects and estimates the spoofing attacks on each state, and mitigates the spoofing on both types of attack by furnishing the corrected clock states to the user. Future work will focus on developing robust estimators under spoofing attacks for non-stationary GPS receivers, which involve nonlinearities in the GPS measurement model. }
	
		\vspace{-0.5cm}
	\bibliographystyle{IEEEtran}	\bibliography{bib_file}
	
\end{document}